\documentclass[%
 reprint,
 amsmath,amssymb,
 aps,
 physrev,
]{revtex4-2}

\usepackage{graphicx}% Include figure files
\usepackage{dcolumn}% Align table columns on decimal point
\usepackage{bm}% bold math
\usepackage{xcolor}
\usepackage{hyperref}% add hypertext capabilities
\hypersetup{colorlinks=true, citecolor=blue, urlcolor=blue, linkcolor=blue}
\usepackage{wasysym}
\usepackage{rotating}

\newcommand{\degree}{${}^\circ$}

\newcommand{\icarus}{Icarus}
\newcommand{\jgr}{J. Geophys. Res.}

\newcommand{\ssr}{Space Sci. Rev.}

\begin{document}

\title{Measurement of the Free Neutron Lifetime using the Neutron Spectrometer on NASA's Lunar Prospector Mission}

\author{Jack T. Wilson}
\email{Jack.Wilson@jhuapl.edu}
\author{David J. Lawrence}%
\author{Patrick N. Peplowski}%
\affiliation{%
 The Johns Hopkins Applied Physics Laboratory,\\
 11101 Johns Hopkins Road,\\
 Laurel, Md. 20723, USA.
}%
\author{Vincent R. Eke}
\author{Jacob A. Kegerreis}
\affiliation{
 Institute for Computational Cosmology,\\ 
 Durham University, South Road, \\
 Durham DH1 3LE, UK.
}%

\date{\today}

\begin{abstract}
We use data from the Lunar Prospector Neutron Spectrometer to make the second space-based measurement of the free neutron lifetime finding $\tau_n=887 \pm 14_\text{stat}{\:^{+7}_{-3\:\text{syst}}}$~s, which is within 1$\sigma$ of the accepted value.  This measurement expands the range of planetary bodies where the neutron lifetime has been quantified from space, and by extending the modeling to account for non-uniform elemental composition, we mitigated a significant source of systematic uncertainty on the previous space-based lifetime measurement.  This modeling moves space-based neutron lifetime measurement towards the ultimate goal of reducing the magnitude of the systematics on a future space-measurement to the level of those seen in laboratory-based experiments.
\end{abstract}

\maketitle

\section{\label{sec:Introduction}Introduction}
The neutron mean lifetime $\tau_n$ is an important parameter for the weak interaction, and more precise knowledge of its value is needed in particle physics, nuclear physics, and cosmology \citep{Wietfeldt2011,Dubbers2011}.  Specifically, $\tau_n$ is a key input to calculations of primordial helium abundance, and the uncertainties in these predictions are presently dominated by those on $\tau_n$ \citep{Cyburt2016}.  Primordial nucleosynthesis is one of the major lines of evidence for the big bang, along with the observed expansion of the universe and cosmic microwave background. Additionally, in combination with beta-decay correlations, $\tau_n$ is used to test the unitarity of the Cabibbo--Kobayashi--Maskawa (CKM) matrix, which is one of the most important low-energy tests of the standard model \citep{Wietfeldt2011}. 

Currently, there are two competing values for $\tau_n$ based on the results of two different classes of laboratory experiments.  The `bottle' experiments involve counting the number of neutrons that survive within a material, magnetic, and/or gravitational trap as a function of time.  The `beam' experiments involve measuring the rate of production of $\beta$-decay products in a neutron beam passing through a trapping region. The average beam measurement ${\tau_n^\textrm{beam} = 888\pm2}$~s differs by about 4$\sigma$ from the more precise ultra-cold trapped neutron average ${\tau_n^\textrm{bottle} = 879.4\pm0.6}$~s.  This discrepancy, which has persisted for 15 years, has become known as the `neutron lifetime puzzle'.  The most likely explanation for the discrepancy is the presence of an unaccounted for systematic error in one, or both, classes of experiment. However, given the direction of the disagreement, a physical explanation is possible where the neutron decays to unobserved particles outside of the standard model with a branching fraction of approximately 1\% \citep[e.g.,][]{Fornal2018,Berezhiani2019}. 

The feasibility of a third technique, measuring $\tau_n$ from space, was recently demonstrated using data taken by NASA's MESSENGER spacecraft during its flybys of Venus and Mercury \citep{Wilson2020}. The opportunity to make a space-based measurement of $\tau_n$ is made possible by the fact that planetary surfaces are constantly bombarded by galactic cosmic rays (GCRs) that collide with atomic nuclei, which leads to the liberation of large numbers of high-energy neutrons.  These neutrons have their energy moderated downwards in subsequent collisions with near-surface nuclei.  Some fraction of the neutrons undergo a sufficiently large number of collisions that they approach thermal equilibrium with the atmosphere or solid surface.  These thermal neutrons typically have velocities on the order of a few km\,s$^{-1}$. Thus, their time of flight between emission and detection by a spacecraft hundreds to thousands of kilometers above the surface is of order $\tau_n$. The neutron flux that ultimately escapes from a planetary body into space is characteristic of that planet's elemental composition on depth scales of order the neutron mean free path, which is typically $\sim$ 10 cm in silicate rocks \citep{Lingenfelter1961}.  As measuring planets' surface compositions often forms a major goal of planetary missions, several neutron spectrometers have been included on such missions to learn about planetary surface composition \citep{Feldman2004,Boynton2004,Goldsten2007,Prettyman2011}.

Although MESSENGER spent several years in orbit around Mercury, uncertainty in that planet's surface composition greatly constrained the accuracy of the resulting measurement, which had to rely on a small amount of data taken during the spacecraft's flyby of Venus \citep{Wilson2020}. In this paper we consider data taken by NASA's Lunar Prospector (LP) mission during its initial highly elliptical orbits of the Moon \citep{Maurice2004}.  Unlike Mercury, the Moon's surface composition is well understood and can be included in the modeling of the detected neutron count rate.  Thus, our results demonstrate the practicality of reducing this major systematic uncertainty in the space-based approach to measuring $\tau_n$. 

\section{\label{sec:Data} LP elliptical orbital data}
LP arrived in orbit at the Moon on 11 January 1998 and spent 18 months gathering data in mapping orbits that were close to circular. The LP neutron spectrometer (NS) was designed to explore the Moon's surface composition with a focus on searching for water ice at the poles and mapping the Moon's elemental composition \citep{Feldman1999}.  The NS consisted of two cylindrical  gas proportional counters 5.7~cm in diameter and 20~cm in length filled with 10 atmospheres of $^3$He.  One of the detectors was covered in 0.63~mm of Cd to shield it from thermal neutrons and the other with 0.63~mm of Sn, which has a similar atomic number to Cd but a much smaller neutron capture cross section and a negligible effect on the number of neutrons detected \citep{Feldman1999}. The spacecraft body was cylindrical with the detectors mounted on the end of a 2.5~m boom perpendicular to the cylinder's long axis to reduce the spacecraft's contribution to the measured signal. The spacecraft was spinning about its long axis at 12 revolutions per minute and the integration time for a single observation was 32~s.  Thus, the response of the detector was an angle-averaged one with respect to rotation about LP's long axis. 

\begin{figure}
\includegraphics[width=\columnwidth]{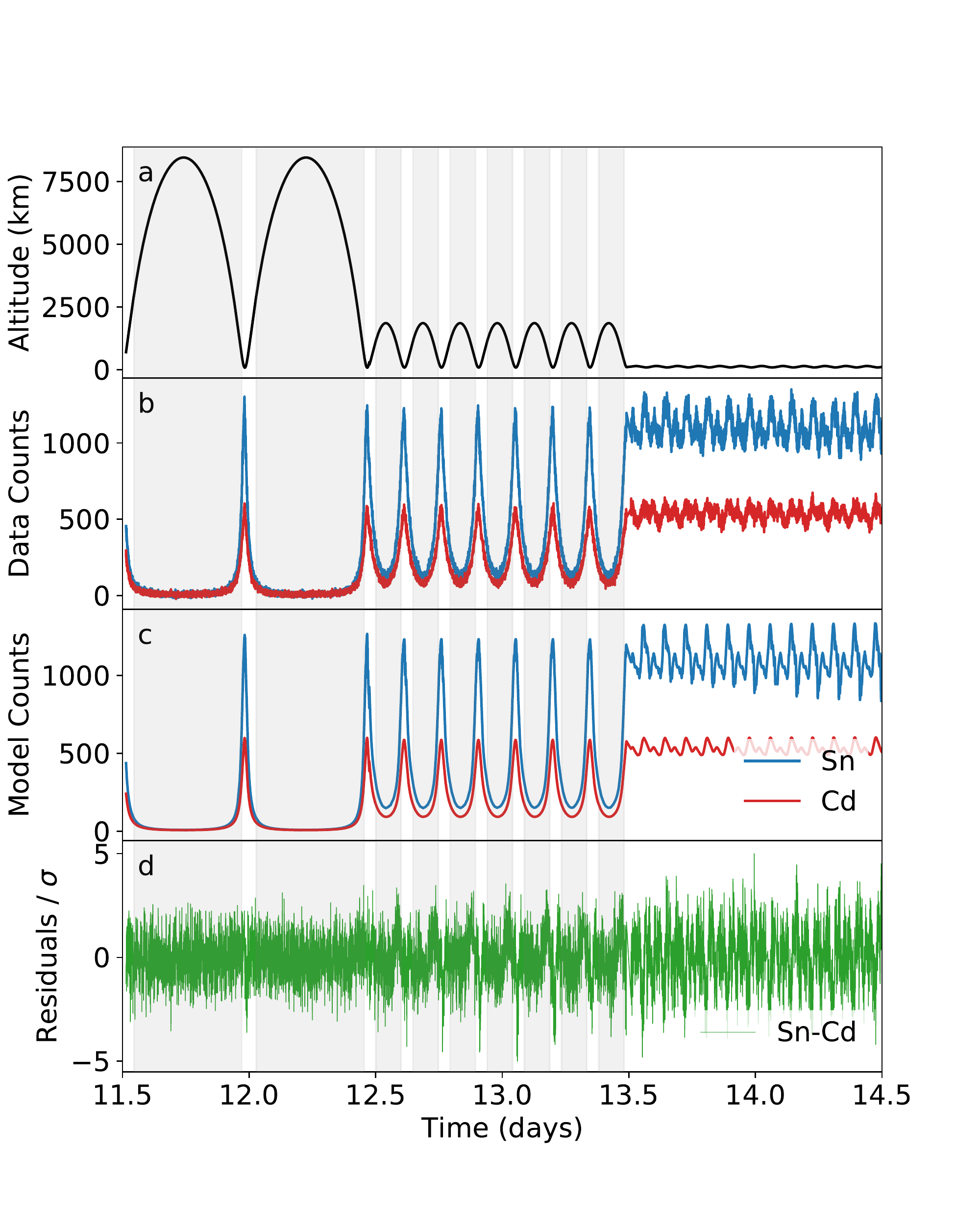}
\caption{\label{fig:cts_zoom}(a) LP altitude during the initial elliptical orbits. (b) The measured count rates in the Cd- and Sn-covered NS detectors. (c) The modeled count rates in the Cd- and Sn- covered detectors as described in section~\ref{sec:Model}. (d) The residuals of the thermal neutron counts (i.e., the difference between the Sn- and Cd- covered detectors) normalized by the uncertainty due to counting statistics. The grey regions show the data used to produce the final result based on the cuts described in section~\ref{sec:Results}. Time on the x-axis is measured from January 1st, 1998.}
\end{figure}

Before LP entered its circular mapping orbits it carried out a set of nine highly elliptical orbits \citep{Maurice2004}. We used the variation in thermal neutron flux  measured during these elliptical orbits, as determined by the difference in the two $^3$He gas proportional counters, to infer $\tau_n$ by minimizing a chi-squared comparison to Monto Carlo simulated fluxes. The spacecraft altitude, along with the measured and modeled neutron count rates for these orbits, is shown in Fig.~\ref{fig:cts_zoom}. 

\begin{figure}
\includegraphics[width=\columnwidth]{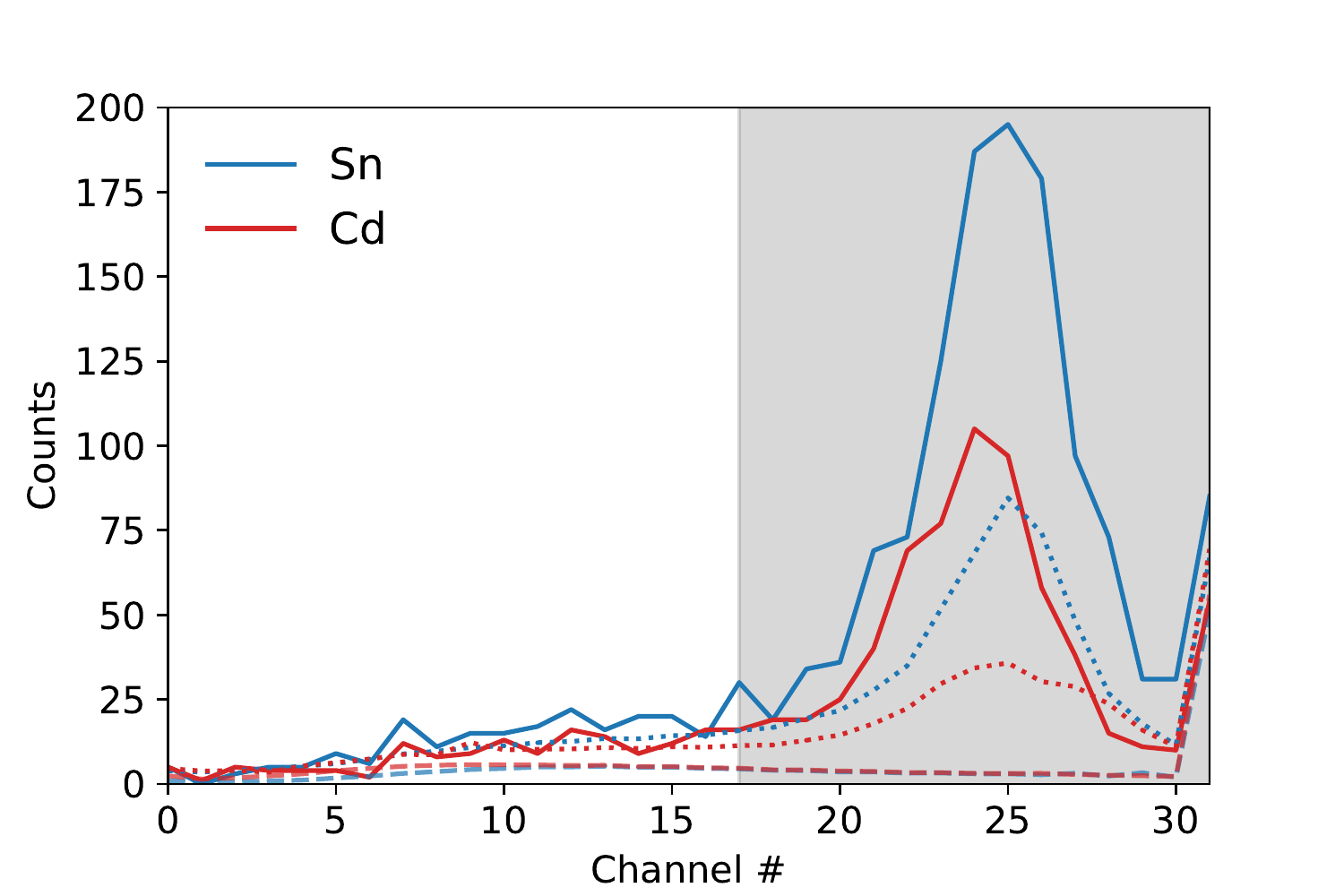}
\caption{\label{fig:cts_spec}Low-altitude 32-s (solid) and mission averaged (dotted) LPNS spectra from both the Cd- and Sn- covered $^3$He gas proportional counters. The grey shaded region shows those channels involved in determining the measured neutron counts. The dashed curves show the high-altitude background, measured during LP's approach to the Moon when the spacecraft altitude was $> 1.5 \times 10^4$~km.}
\end{figure}

The observations of each $^3$He detector were recorded as 32-channel pulse height spectra every 32~s (Fig.~\ref{fig:cts_spec}). The measured spectra include the 764~keV energy deposition peak from the ${n + {}^3\textrm{He} \ensuremath{\rightarrow} {}^3\textrm{H} + {}^1\textrm{H}}$ reaction, which is the signal of interest. Due to the spacecraft design the continuum background originating from the interaction of GCRs with the spacecraft and detector is small (Fig.~\ref{fig:cts_spec}).  The conversion of recorded spectra to count rates involved removing this background and then summing over the channels that measure the neutron absorption peak. Due to the short duration of the elliptical phase of the mission, GCR and gain variation corrections were not necessary, thus the data reduction is much simplified compared with analyses of the full LPNS mission dataset \citep[e.g.,][]{Maurice2004}.

The mean LPNS background spectrum was estimated by averaging the spectra taken during the spacecraft's initial approach to the Moon when its altitude was greater than $1.5 \times 10^4$~km. Background-subtracted spectra were then produced by subtracting a solid-angle-scaled high-altitude background $b^\prime$ from each individual spectrum in the time series, i.e.,
\begin{align*} 
b^\prime &=  fb, \\ 
f &= \frac{1 + \cos(\theta)}{2}, \\
\theta &= \sin^{-1}\left(\frac{R_{\leftmoon}}{R_{\leftmoon} + h}\right),
\end{align*}
where $b$ is the high-altitude background spectrum, $R_{\leftmoon}$ is the mean radius of the Moon and $h$ the spacecraft's altitude above the Moon's surface. This solid angle scaling is a consequence of the fact that the background is largely due to the interaction of GCRs with the spacecraft, which are obscured by the Moon when the spacecraft is at low altitudes.  To calculate the neutron count rate, the background-subtracted spectra were summed over the channels containing the neutron-absorption peak (shown with the grey shaded region in Fig.~\ref{fig:cts_spec}) before being divided by the observation period.  There is a small shift in the location of the peak in the pulse height spectra associated with temperature changes in the detectors, however the summation region is sufficiently large that these variations have no effect on the results.  Uncertainties are those resulting from the Poisson statistics of the observed spectra. Data from both of the $^3$He detectors are shown in Fig.~\ref{fig:cts_zoom}(b).

\section{\label{sec:Model}Neutron production and transport modeling}

As in \citet{Wilson2020} we made use of a comparison between the measured data and the result of models of neutron production, transport, and detection to estimate $\tau_n$.  The modeled count rates were determined using a combination of three separate calculations. First, we used the particle transport code MCNPX \citep{Pelowitz2005} to model the neutron production following GCR interaction with the Moon's surface and the flux due to the escape of these neutrons for the set of compositions shown in Table~\ref{tab:comps}, with the Moon taken to be a uniform sphere of radius 1738~km. The GCR flux was assumed to be isotropic, with an energy distribution described by the parameterization of \citet{Masarik1996} with solar modulation parameter equal to 550~MV. Due to the low energy of the incident protons (E~$\lesssim$~10~MeV) the contribution of the solar wind to neutron production is negligible outside of brief solar energetic particle events \citep{McKinney2006}, none of which occurred during LP's orbital insertion.

\begin{figure}
\includegraphics[width=\columnwidth]{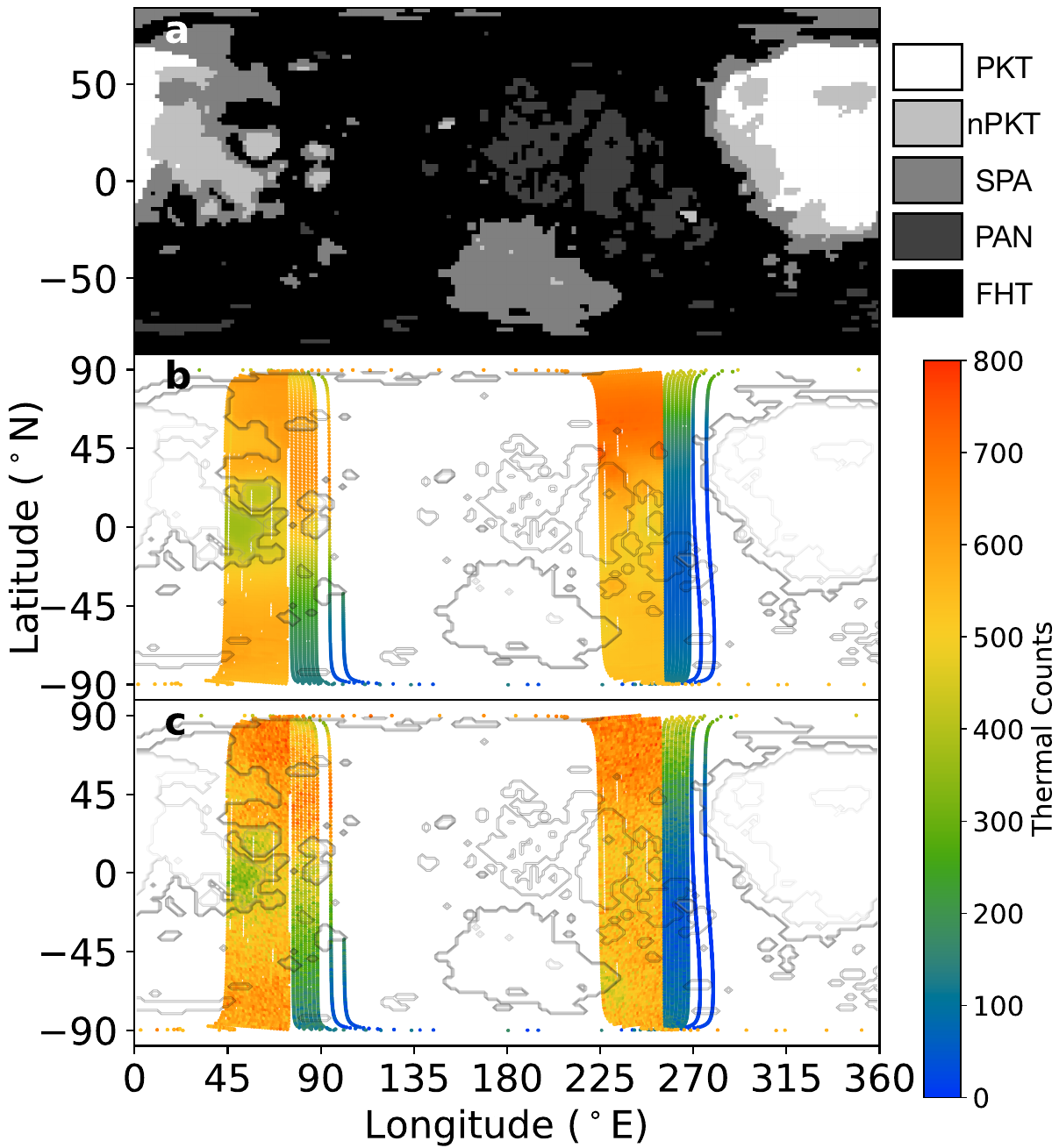}
\caption{\label{fig:regions}(a) The five regions with distinct compositions included in the neutron count-rate models: Procellarum KREEP terrane (PKT), South Pole Aitken Terrane (SPAT),  Feldspathic Highland terrane (FHT),  Pure Anorthosite terrane (PAN), and non-PKT maria (nPKT). (b) The model thermal neutron counts, i.e., the difference between the Sn-covered and Cd-covered detectors. (c) The measured thermal detector counts. The grey contours in (b) and (c) show the regions defined in (a).}
\end{figure}

In the second step of the modeling, an extended version of the formalism of \citet{Feldman1989} was used to calculate the flux at the spacecraft altitude by analytically extending the surface fluxes calculated in the previous step. Here, we have modified the approach to include the effect of variations in the planet's surface composition on the neutron flux. The Moon's surface is conventionally taken to consist of three major compositional terranes: the Procellarum KREEP terrane (PKT), a high-Th, magnesium and iron rich province on the lunar nearside; the South Pole Aitken (SPA) terrane, an impact feature with intermediate-Th abundances on the lunar farside that may include material from the upper mantle and lower crust; and the Feldspathic Highland terrane (FHT), an anorthositic region exposing the result of early lunar crustal differentiation \citep{Jolliff2000}. To capture the range of thermal neutron variation across the lunar surface, we add two supplemental regions: the nearly Pure Anorthosite (PAN) terrane\citep{Peplowski2016}; and the non-PKT maria (nPKT), a mafic but less Th-rich companion to the PKT. The PAN region has a thermal neutron flux significantly higher than the surrounding regions due to its deficit in rare earth elements (REE) with large neutron absorption cross sections. We used the LP low-altitude (30~km average) thermal neutron data to define its extent. The nPKT has a thermal neutron flux intermediate between the PKT and highlands due to its high Fe but low REE composition and its boundaries were drawn based on LP low-altitude Th gamma-ray line and fast neutron data, using Th as a proxy for REE \citep{Elphic2000} and fast neutrons as representative of Fe. The extent of each of the regions is shown in Fig.~\ref{fig:regions}. 

The compositions of each of the regions to be used in the models were based on our current best understanding of lunar near-surface bulk composition as described in several studies of LP NS and GRS data \citep{Elphic2000,Prettyman2006,Peplowski2015,Lawrence2015}. The major and radioactive element compositions of each region were determined by averaging, over each region, the elemental abundances derived in \citet{Prettyman2006}, which analyzed LP GRS spectra taken during the initial circular mapping orbit when the spacecraft had an average altitude of 100~km.  For the rare earth elements Sm and Gd we used the abundances derived in \citet{Elphic2000}, which are based on thermal neutron measurements taken during the low altitude phase of the LP mission when the spacecraft had an average altitude around 30~km. The H abundances were derived from \citet{Lawrence2015}.  Although these compositions provide our best estimate of the elemental abundances of the lunar surface  the elemental abundances determined using this method do not yield macroscopic neutron absorption cross sections $\Sigma_a$ that match the average values of $\Sigma_a$ measured for each region ~\citep{Peplowski2015}. As our measurement is determined by the thermal neutron flux, having lunar soils with the correct $\Sigma_a$ in our models is important. Therefore, the final step in setting the soil compositions was to alter the elemental abundances so that $\Sigma_a$ in the simulated soils matched the measured values. We chose to modify the relative abundances of the most abundant element, oxygen, while maintaining the relative abundances between each of the other elments. The size of these final modifications to the oxygen abundances is small, ranging from a 0.3\% decrease for the FAN composition to a 5\% decrease for PKT soil. The compositions used in this study are shown in Table~\ref{tab:comps}.

The non-uniformity of the Moon's surface composition was included in the modeling of neutron transport by modifying the equations developed in \citet{Feldman1989} such that the flux at the detectors at an altitude $R$,  latitude $\lambda_R$, and longitude $\varphi_R$ is related to the surface flux at latitude $\lambda$, and longitude $\varphi$ by 
\begin{equation*}
	\Phi_R\left[\lambda_R,\varphi_R \right](K_R,\mu_R,\phi_R) = \left(\frac{K_R}{K}\right)\Phi[\lambda,\varphi](K,\mu,\phi)e^{-\frac{\Delta t_R}{\tau_n}},
\end{equation*}
where $K$ is the kinetic energy of the neutron at emission, $K_R=K - V(R-R_{\leftmoon})$ is the kinetic energy at $R$, with $V = \frac{GMm}{R_{\leftmoon}}$, $M$ the mass of the Moon, $m$ the neutron mass, $\mu$ the cosine of the angle of emission with respect to the local zenith $\theta$, $\mu_R = \sqrt{1-(R_{\leftmoon}/R)^2(K/K_R)(1-\mu^2)}$, and $\Delta t_R$ is the time for transit for a neutron travelling from the surface to an altitude $R$ \citep{Feldman1989}, which is given by the expression
\begin{multline*}
\Delta t_R = \frac{R_{\leftmoon}(m/2V)^{\frac{1}{2}}}{2(1-K/V)^{\frac{3}{2}}}\left(2\mu\left(1-\frac{K}{V}\right)^{\frac{1}{2}}\left(\frac{K}{V}\right)^{\frac{1}{2}} \times \right. \\
 \left(1 - \left|\frac{\tan{\theta}}{\tan{\theta_R}}\right|\right) + \sin^{-1}\left(\frac{B}{(A^2+B^2)^{\frac{1}{2}}} \right) \\
\left. + \sin^{-1}\left(\frac{1 - 2K_R/V_R}{(A^2+B^2)^{\frac{1}{2}}} \right) \right)
\end{multline*}
with
\begin{multline*}
	A = \sqrt{4\left(\frac{K}{V}\right)\left(1 - \frac{K}{R}\right)\mu^2}, \quad B = \left(\frac{2K}{V} - 1\right) \\ \text{for}\quad \frac{K}{V} < 1.
\end{multline*}
A similar expression exists for $K/V > 1$, see \citet{Feldman1989} for more details and a derivation of the above equation. $\Phi[\lambda,\varphi](K,\mu,\phi)$ is the flux at the surface of the Moon, which is found by looking up the appropriate MCNPX output flux for the model composition at latitude $\lambda$, and longitude $\varphi$. The location of emission of a neutron is uniquely determined by the location and velocity at detection by Kepler's laws and spherical trigonometry such that
\begin{equation*}
	\lambda = \sin^{-1}\left(\sin(\lambda_R)\cos(\delta) + \cos(\lambda_R)\sin(\delta)cos(\pi-\phi_R)\right)
\end{equation*}
and
\begin{equation*}
	\varphi = \varphi_R + \tan^{-1}\left(\frac{\sin(\pi-\phi_R)\sin(\delta)\cos(\lambda_R)}{\cos(\delta) - \sin(\lambda_R)\sin(\lambda)} \right),
\end{equation*}
where
$\delta$ is the difference between the true anomaly at neutron detection $\nu_0$ and emission $\nu_1$.  The true anomaly can be calculated from the semi-major axis
\begin{equation*}
	a = -\frac{VR}{2(K_V)}
\end{equation*}
and ellipticity
\begin{equation*}
	e = \sqrt{1 + 4\frac{K}{V}\left(\frac{K}{V} - 1\right)(1 - \mu^2)}
\end{equation*}
via
\begin{equation*}
	\nu_0 = \cos^{-1}\left(\frac{a(1-e^2) - R}{eR} \right).
\end{equation*}

\begin{figure}
\includegraphics[width=\columnwidth]{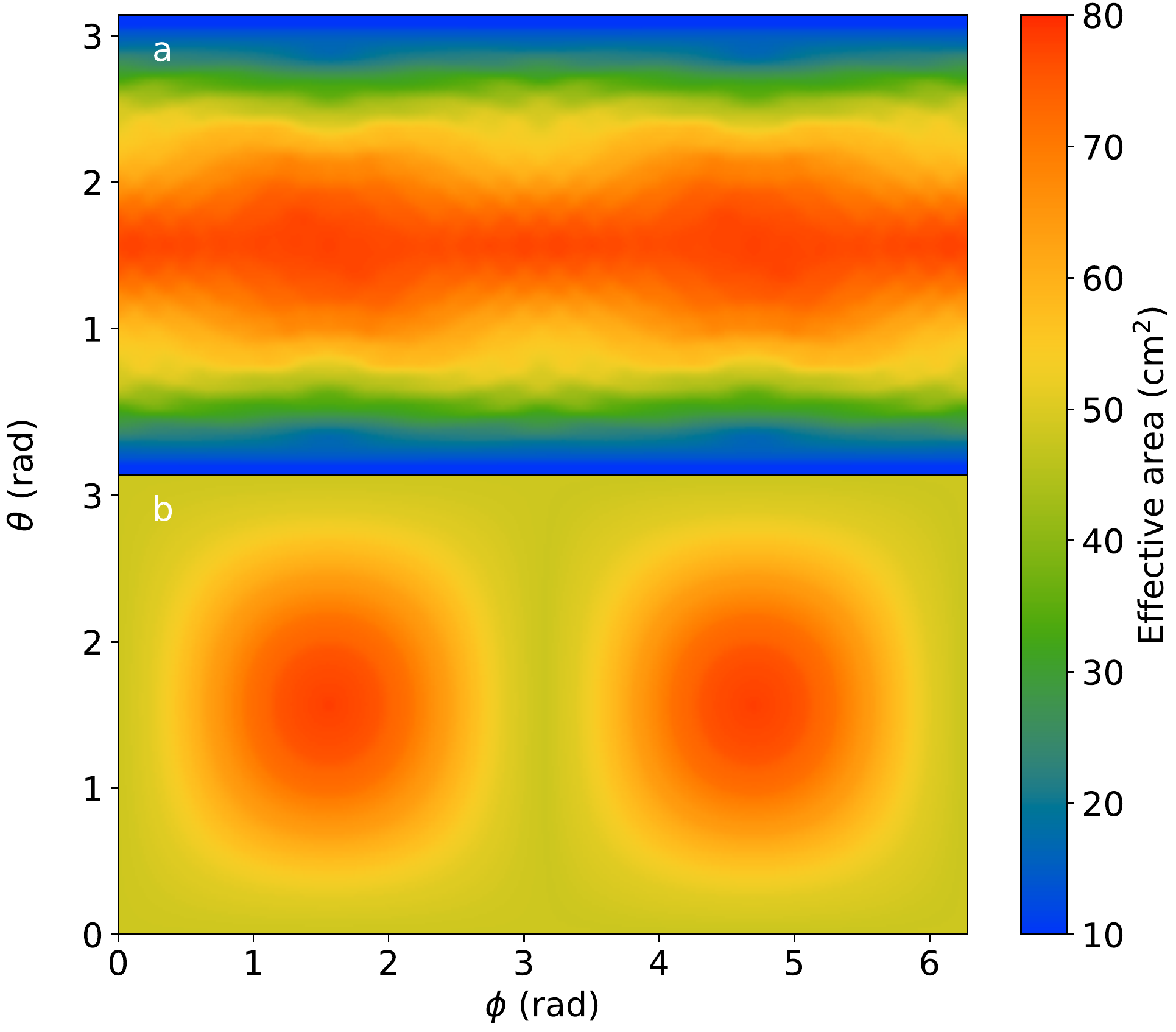}
\caption{\label{fig:response}(a) The result of the MCNPX-based efficiency calculation for a Sn-covered ${}^3$He tube to $10^{-9}$~MeV neutrons. (b) An example of angle-averaged detector response used in the count rate simulation, based on the instantaneous response shown in (a).  The coordinates are conventional spherical polars where the spacecraft's rotation axis coincides with the y-axis and the NS detector's long axis coincides with the z-axis.}
\end{figure}

\begin{figure}
\includegraphics[width=\columnwidth]{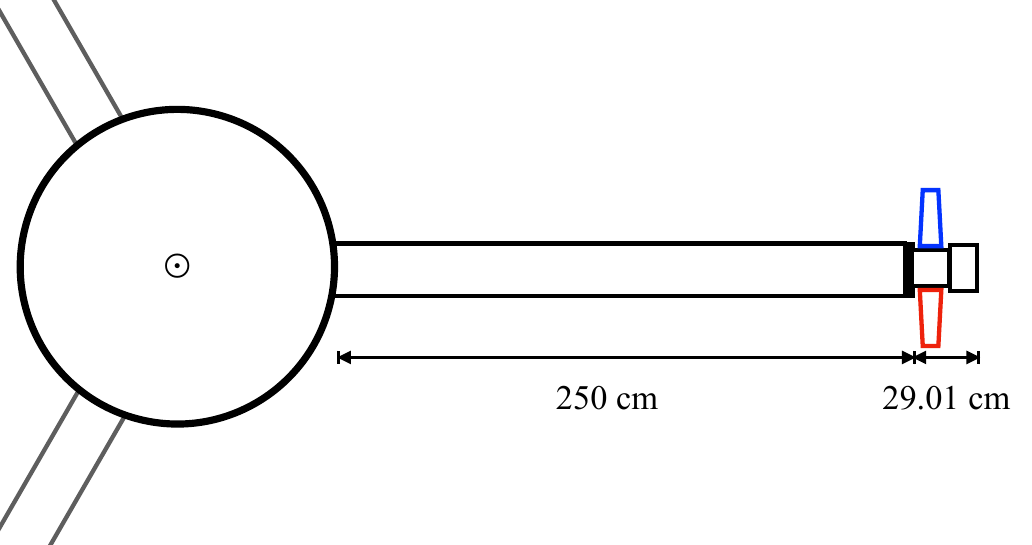}
\caption{\label{fig:LP} A cartoon of the LP spacecraft and NS looking along the spacecraft rotation axis. The locations of the Sn-covered and Cd-covered tubes are shown in red and blue, respectively. The axis of rotation, pointing out of the page, is indicated with the circled dot.}
\end{figure}

The third step in the generation of model count rates was the creation of an MCNPX \citep{Pelowitz2005} model of the NS detectors and the calculation of their response to neutrons that are incident with different momenta. Due to the small size of the spacecraft and the location of the detectors on the end of a 2.5~m boom (Fig.~\ref{fig:LP}), a full simulation of the spacecraft was not necessary.  Instead the small spacecraft-originating background was removed from the data as described in section~\ref{sec:Data}.  The NS detectors were modeled in MCNPX  as cylinders 20~cm long and 5.7~cm in diameter filled with $^3$He with a density of $1.2\times 10^{-3}$~g\,cm$^{-3}$.  Two detectors were simulated, one covered in each of 0.63~mm of Cd and Sn with respective densities of 8.67~g\,cm$^{-3}$ and 7.29~g\,cm$^{-3}$. These detectors were illuminated with plane wave neutrons with energies spanning the range to which the detectors are sensitive, between $10^{-9}$~MeV and $10^{-2}$~MeV, and a range of incidence angles covering the sphere.  To calculate the effective area $A$ of the detectors the total number of neutrons absorbed in the active region of the detectors was tallied, where the active region was defined as between 4.975~cm from one end and 3.705~cm from the other, and multiplied by the number of primary events per unit area of the simulated neutron source.  This estimate of the active region is based on recent calibration measurements of similar $^3$He sensors being used for NASA's Psyche mission \citep{Peplowski2020}. The result of the simulation of the Sn-covered tube following illumination with $10^{-9}$~MeV neutrons is shown in Fig.~\ref{fig:response}(a).  As LP rotates approximately six times during a single observation period of the NS, the effective detector response was taken to be the instantaneous detector response, calculated in MCNPX, averaged over rotation about the spacecraft's rotation axis.  This is shown in Fig.~\ref{fig:response}(b). A cartoon of the spacecraft and detectors is shown in Fig~\ref{fig:LP}.

The final step in the generation of model count rates is to combine the models of surface neutron flux, neutron transport and detector effective area by integrating over neutron detection angle and energy such that 
\begin{multline*}
	J(\lambda_R,\varphi_R) = \int_0^\infty dK_R \int_{-1}^1 d\mu_R \int_0^{2\pi} d\phi_R  \left(\frac{-v\cdot\hat{n}}{v_n} \right) \times \\ 
	A(K_R^\prime,\mu_R^\prime,\phi_R^\prime) \Phi_R\left[\lambda_R,\varphi_R \right](K_R,\mu_R,\phi_R),
\end{multline*}
where $v$ is the spacecraft velocity, $v_n$ the neutron velocity, and $\hat{n}$ the detector normal. The velocities are in the Moon's reference frame and primed quantities are those in the frame of the detector.  The integration over neutron flux at the detector includes the contribution of both upward going neutrons coming directly from the lunar surface and downward going neutrons on elliptical trajectories that are returning to the Moon's surface.  This effectively integrates over neutrons originating from across the lunar surface.  

The results of these calculations for both the Cd-covered and Sn-covered $^3$He detectors is shown in Fig.~\ref{fig:cts_zoom}(c). Fig.~\ref{fig:regions}(b) shows the modeled thermal neutron counts, which is the difference between modeled Sn-covered and Cd-covered detector counts.  It is the thermal neutron counts that will be used to determine $\tau_n$ due to their greater sensitivity to that parameter, which results from the lower velocity of these neutrons yielding a greater time of flight.  

During LP's mapping phase, the spacecraft was spin stabilized about an axis normal to the plane of the ecliptic. However, in the early elliptical orbits this axis varied and was typically around 60\degree\ to the elliptic. The spacecraft's orientation was recorded before and after each maneuver and this measured orientation was included in the modeling.

\begin{figure}
\includegraphics[width=\columnwidth]{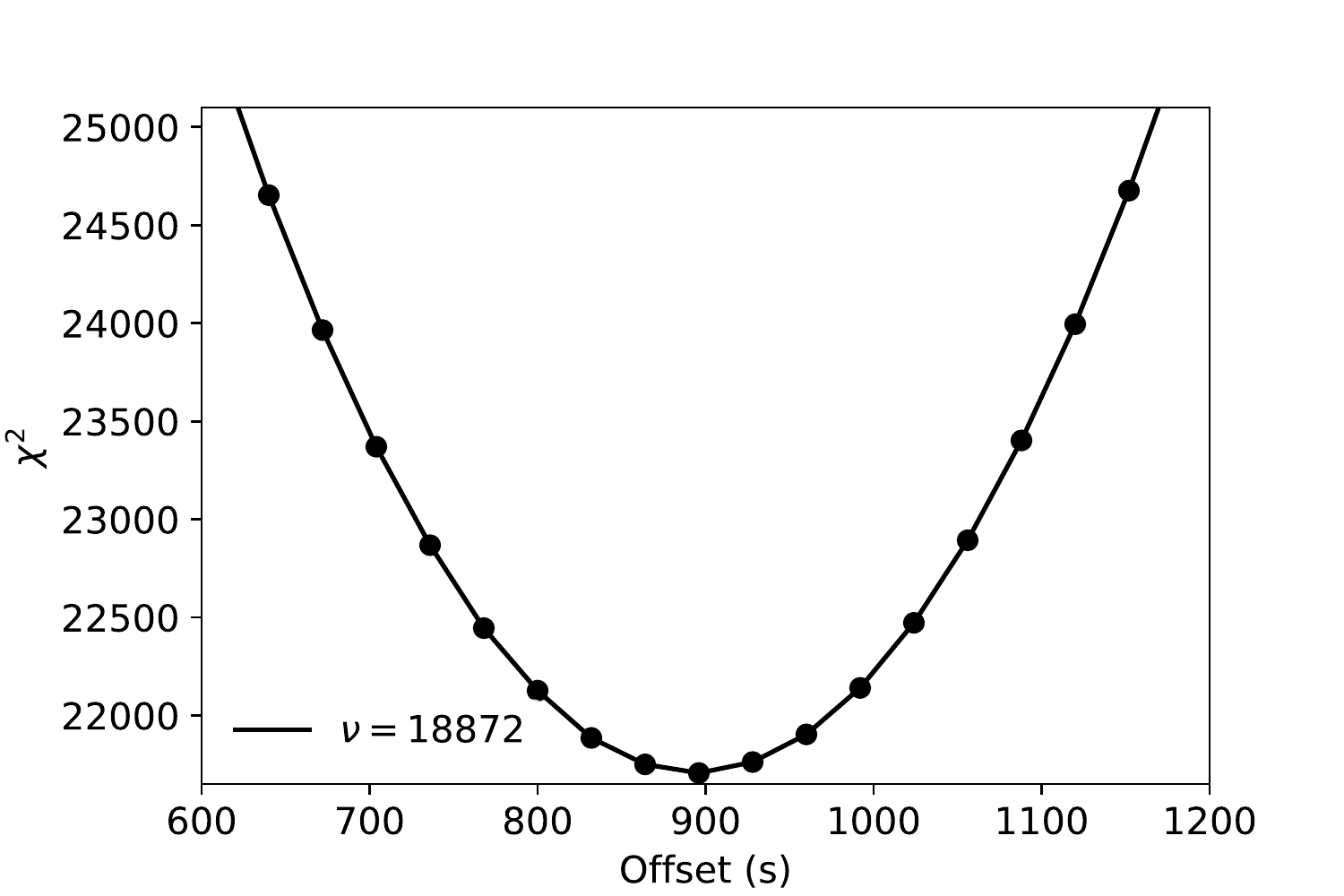}
\caption{\label{fig:offset}$\chi^2$ statistic based on a comparison between the GRS counts and solid-angle subtended by the Moon, with a linear scaling applied to minize the misfit, for different temporal offsets applied to the spacecraft ephemerides. The number of degrees of freedom $\nu$ is shown in the figure.}
\end{figure}

As the early elliptical orbits used in this study were not part of the science phase of the LP mission they have been little used and not extensively verified.  We observed that there was a temporal offset between certain spacecraft ephemerides (importantly altitude and velocity) and the detected counts in the NS and the Gamma-Ray Spectrometer (GRS). That is, the times of closest approach were not coincident with maxima in the rate of gamma ray detection. To find the size of this temporal offset we tried shifting the spacecraft ephemerides in time by increments of the NS integration period of 32~s. A $\chi^2$ comparison of solid-angle subtended by the Moon, following various time offsets, with GRS counts integrated over all energies is shown in Fig.~\ref{fig:offset}, where the solid-angle has been scaled linearly to fit the GRS counts.  The magnitude of this offset was found to be $900.0 \pm 0.5$~s, which was estimated by fitting a quadratic to the points shown in Fig.~\ref{fig:offset}. A comparison with the GRS data is preferable to one with the NS measurements due to the Doppler effects in the neutron data resulting from the particles' velocities being comparable with that of the spacecraft.  We accommodated the offset in ephemerides by translating the altitude and velocity parameters in time with respect to the NS-measured counts. It will be shown in Section~\ref{sec:Results} that the final agreement between the data and model is good, which confirms the success of this correction.  Additionally, the spacecraft ephemerides for the first half of the first elliptical orbit were missing so these data are not included in the modeling.

Similar to \citet{Wilson2020}, the basis of the measurement technique is a comparison of the NS-measured data with models of the constructed count rate assuming different values of $\tau_n$.  However, unlike that earlier study, here we will use only the change in detected neutron counts with altitude to measure the $\tau_n$, not the absolute count rate.  In this approach each model is normalized directly to the data with a single multiplicative factor chosen to minimize the misfit between the model and data. This change mitigates the largest source of systematic uncertainty in the previous measurement, which was associated with errors in the normalization.  Direct normalization of the models to the data is made practical by the larger amount of data taken by LP during its elliptical orbits compared with MESSENGER's flyby of Venus. 

\section{\label{sec:Results}Results}
We used the thermal neutron data (i.e., the difference between the Sn- and Cd-covered NS detectors) to make the measurement of $\tau_n$, due to the larger changes in the thermal neutron count rate compared with those in either the Cd-covered or Sn-covered detectors, for given changes in $\tau_n$. Focussing on the region of measurement-space that is most sensitive to the signal of interest is important when attempting to measure $\tau_n$ at the Moon, due to its relatively low mass and consequently small binding energy (0.029 eV compared with 0.7 eV at Venus and a typical thermal neutron velocity of 0.1~eV) leading to a less steep change in the thermal neutron flux with altitude around the Moon than at Venus \citep{Hess1961}.

\begin{figure}
 \includegraphics[width=\columnwidth]{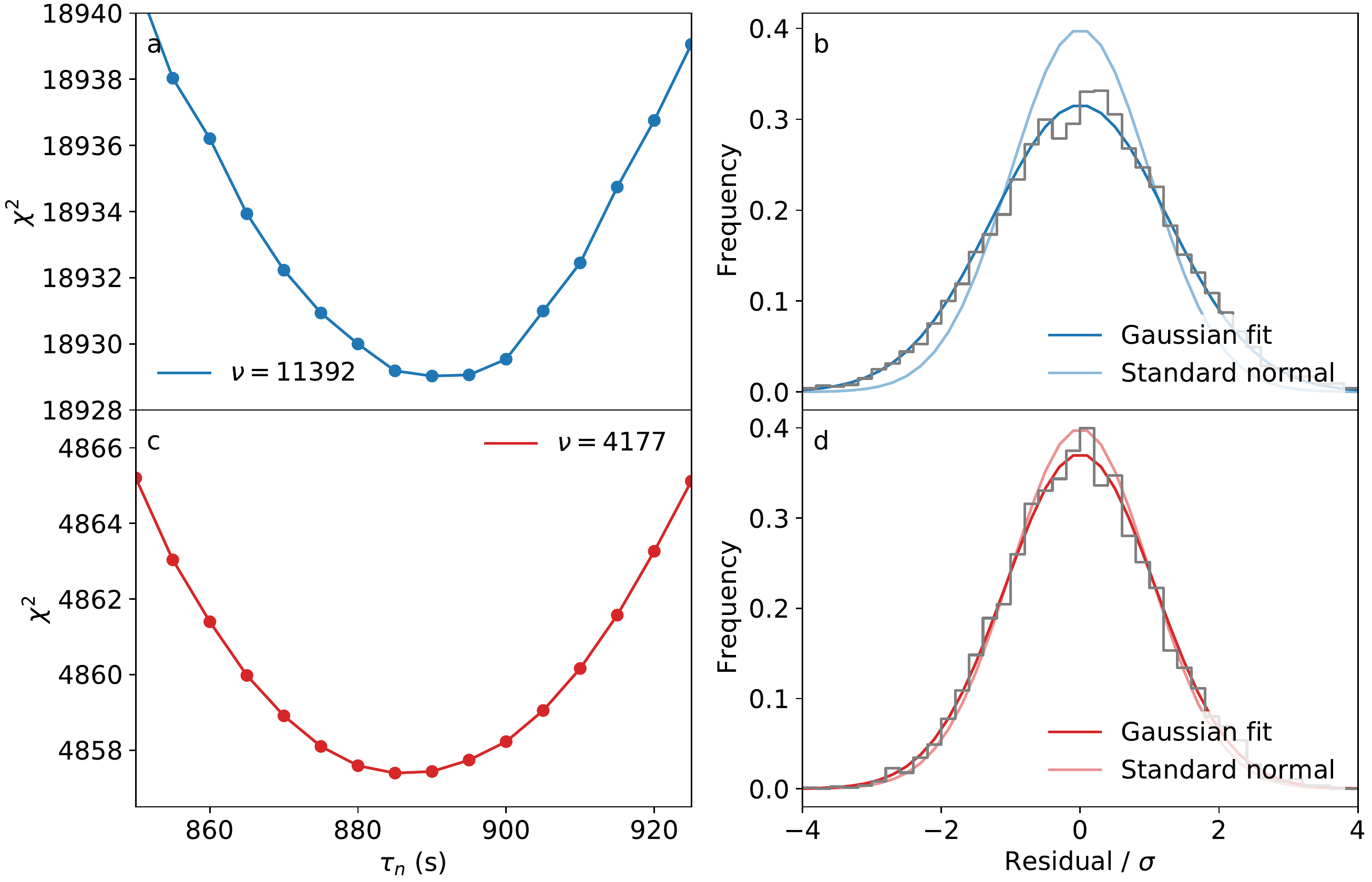}
\caption{\label{fig:CS}(a) $\chi^2$ comparison of the thermal count model and data with differing $\tau_n$ values using all of the observations shown in Fig.~\ref{fig:cts_zoom}. The number of degrees of freedom $\nu$ is shown in the figure. (b) Histogram of the residuals of the best-fit model normalized by the uncertainty due to counting statistics.  (c) and (d) are similar to (a) and (b), respectively, but make use only of the observations taken at longitudes greater than 180\degree during the nine initial highly elliptical orbits.}
\end{figure}

A $\chi^2$ comparison of modeled thermal counts, derived assuming various lifetimes, with the measured data using the uncertainties due to counting statistics is shown in Fig.~\ref{fig:CS}(a).  This comparison includes all of the observations shown in Fig.~\ref{fig:cts_zoom}, including those taken after the initial nine highly elliptical orbits when the spacecraft was in a near circular orbit.  Although qualitative agreement between the model and data can be seen in Figs.~\ref{fig:cts_zoom}~\&~\ref{fig:regions} it is clear from the value of $\chi^2$ shown in Fig.~\ref{fig:CS}(a) and the distribution of residuals in Fig.~\ref{fig:CS}(b) that even the best-fitting model does not adequately describe the data. Therefore, the value of $\tau_n$ obtained from the analysis in Fig.~\ref{fig:CS}(a) is subject to significant systematic errors.  

To avoid some of these systematic uncertainties we confine our analysis to those observations taken where LP's longitude was greater than 180\degree\ and before the near-circular mapping orbit was reached (the grey regions in Fig~\ref{fig:cts_zoom} show the extent of this restricted data set).  With the first of these constraints we removed the eastern lunar maria from the study as the residuals in this region were anomalously large, which suggests that the complex composition of this region has not been fully incorporated into the model.  The effect of the second constraint was to increase the average altitude of observation, which reduces the need to precisely and accurately describe the lunar surface on small scales in the model. The results of these data cuts on the analysis are shown in Fig.~\ref{fig:CS}(c,d).  Although the statistical uncertainty in the result was increased by the reduction in data volume, the  new residuals are close to the expected standard normal distribution. However, the minimum value of $\chi^2$ is still not entirely consistent with a completely well-fitting model given the expected statistical uncertainties on the measured data.  This implies that there exist additional small systematic uncertainties present in the analysis.

\begin{figure}
 \includegraphics[width=\columnwidth]{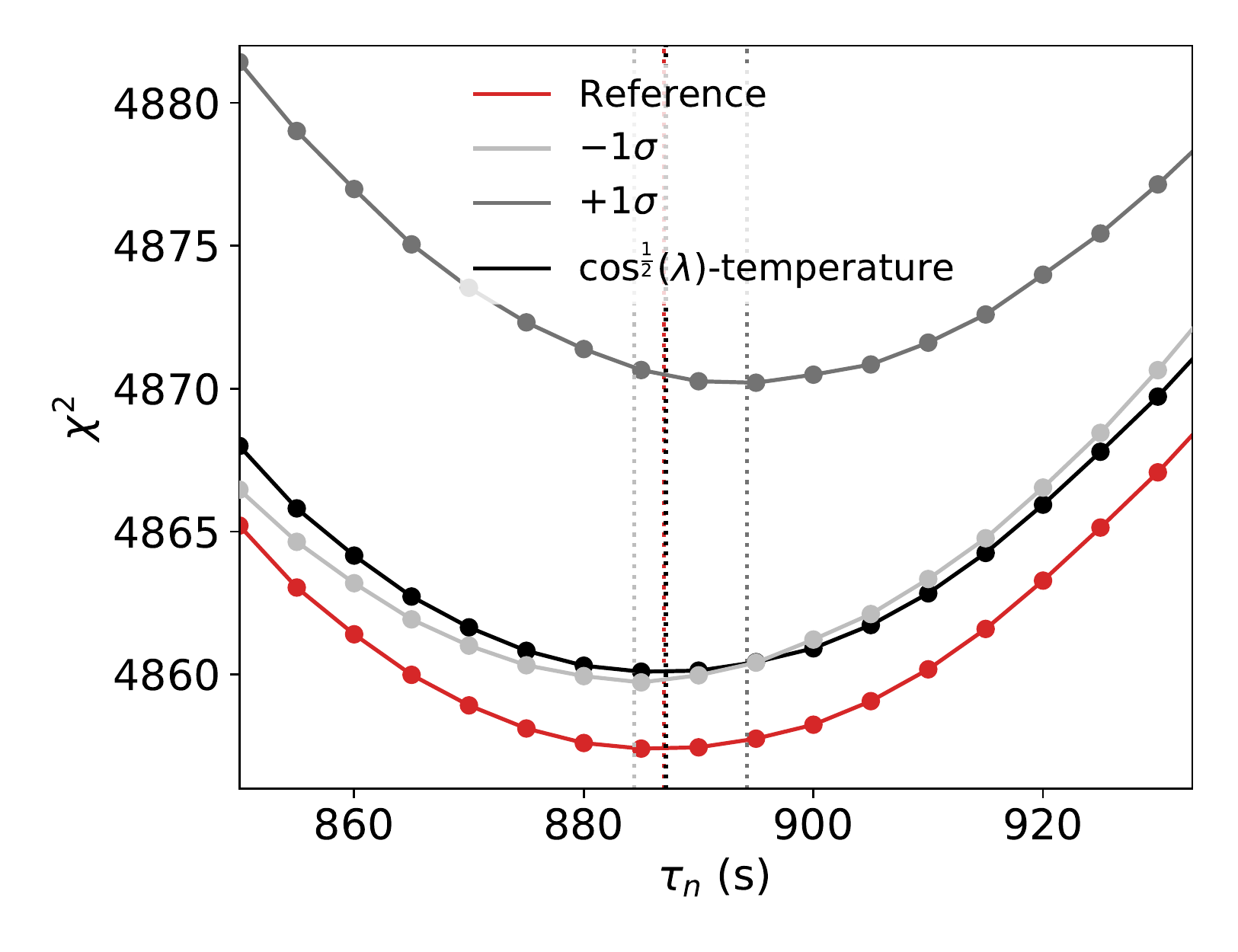}
\caption{\label{fig:sys_comp}$\chi^2$ comparison of the modelled restricted thermal neutron count and data with differing $\tau_n$ values. The red curve shows a comparison with the reference model as in Fig.~\ref{fig:CS}c.  The grey curves are based on models using perturbations from the reference composition.  The black curve compares the data with a model including a latitude-dependent perturbation to the count rates intended to simulate the effect of temperature variation.  The dotted lines indicate the locations of the minima of each of the curves.}
\end{figure}

We expect the largest source of systematic error to result from uncertainties in the compositions used to generate the modeled counts. Based on the elemental covariances reported in \citet{Prettyman2006} we explore the magnitude of this systematic by modifying the compositions used for each of the regions described in section~\ref{sec:Model}.  Specifically, we perturbed the Fe abundance by one standard deviation, in both the positive and negative directions, and changed the abundances for each of the other elements by the amount expected given these changes in Fe concentrations and the reported covariances.  Fe was chosen as variations in that element are the dominant cause of changes in $\Sigma_a$ across the lunar surface \citep{Elphic2000}. We consider this to represent a `$\pm$1$\sigma$' perturbation of the reference compositions.  The elemental abundances of these perturbed compositions are given in Table~\ref{tab:comps2}. As expected both of these models provide a less good fit to the data with the minimum value of $\chi^2$ being greater than that calculated for the reference composition (Fig.~\ref{fig:sys_comp}). A comparison of the $+1\sigma$ models with the restricted data reveals that the best-fit lifetime is approximately 7~s greater than the reference model, for the negative perturbation the best-fit lifetime is around 3~s less than the reference case.  This analysis provides an estimate of the systematic uncertainty in our analysis that results from uncertainties in the lunar composition that is less than 10~s.  Additional systematic uncertainties related to the definition of the regions used in the model exist. Exploration of such systematics is a critical outstanding task for understanding whether a spaced-based measurement of $\tau_n$ could ever be competitive with laboratory measurements. However, given the physical motivation for our selections of these regions and the large parameter space from which they might be chosen, developing this task is left for a detailed study of the potential systematics on a dedicated mission to measure $\tau_n$.

An additional source of systematic error is the effect of variation in lunar temperature with time of day and latitude, which was not included in the modeling but will affect the thermal neutron flux above the Moon. Substantial diurnal variation in thermal neutron flux due to temperature changes is not expected as the diurnal temperature variation at a depth of 20~cm, which is approximately the mean depth of origin of neutrons, is less than 30 K \citep{Vasavada1999,Little2003}.  However, latitudinal variations in temperature are more significant.  The mean surface temperature on the Moon, which is similar to the temperature beneath the diurnal thermal wave envelope, varies between approximately 70~K at the poles and 210~K at the equator \citep{Williams2017} as $\cos^{\frac{1}{2}}\lambda$. \citet{Lawrence2006} showed that similar variations in temperature gives rise to a 2\% variation in thermal neutron count rate when considering LP above a surface of uniform composition and temperature. We estimated the magnitude of the effect on $\tau_n$ of latitudinal variation in temperature by superimposing a 2\% equator-to-pole variation, of $\cos^{\frac{1}{2}}\lambda$ functional form, on the modeled thermal neutron count rate.  The effect of this modification on $\chi^2$ is illustrated in Figure~\ref{fig:sys_comp}.  Fitting a parabola to the reference and surface-temperature adjusted curves in the figure implies a difference in $\tau_n$ of 0.2~s, which is negligible compared with both the statistical errors and the composition-derived systematic described above. The increase in $\chi^2$ over the reference model suggests that this simplistic inclusion of the effect of temperature variation fails to accurately capture the true variation in count rates due to surface temperature. For a precise measurement of $\tau_n$ using data from a dedicated mission these diurnal and latitudinal temperature variations could be included in the model in a similar manner to the current implementation of variable composition. The remaining systematic uncertainty would then be the result of the limits of our knowledge of the lunar subsurface temperature. 

As noted in Section~\ref{sec:Model} a temporal correction to the ephemerides was required.  The 1-$\sigma$ uncertainty on this correction, determined by the temporal offset with $\chi^2$ equal to one greater than the minimum value, is 0.5~s, much less than one observation period of 32~s as shown in Fig.~\ref{fig:offset}.  This ephemerides offset is unusually large and its presence is a consequence of the data not being part of the science phase of the mission, so not previously being well validated. We would therefore not expect similar ephemerides-derived systematics to be present on any future mission.  Ordinarily, uncertainties in spacecraft position are on the order of meters \citep{Lawrence2020} rather than the hundreds of meters travelled by LP during a 0.5-s interval.

% % % An attempt was made to explore the size of this systematic uncertainty by using alternative compositions for the regions defined in Fig.~\ref{fig:regions}(a) based on those representing the Apollo and Luna landing sites (the set in Table~1 of \citet{Lawrence2006}). These landing site compositions are obviously appropriate for certain locations on the Moon's surface and are broadly similar to the compositions in Table~\ref{tab:comps}. However, all of the alternative models produced $\chi^2$ values that were more than 10 larger than the best-fitting model in Fig.~\ref{fig:CS}(c), which further suggests that the remaining uncertainty in the estimate of $\tau_n$ due to compositional uncertainty is small.

For a space-based method similar to that described here to become competitive with current and near-future laboratory-based measurements, the above sources of systematic uncertainty along with potentially significant contributions due to uncertainties in the instrument response function and the Monte Carlo modeling, including nuclear cross-section uncertainties, will need to be corrected for or mitigated \citep{Lawrence2020}.

% \begin{table}
% \caption{\label{tab:sys}Summary of systematic uncertainties associated with the measurement of $\tau_n$}
% \begin{ruledtabular}
% \begin{tabular}{lc}
% Source of uncertainty &Uncertainty (s)\\
% \hline
% Surface composition & 10\\
% Subsurface temperature & 0.2 s \\
% Ephemerides offset & $<$ 5\\\hline
% Total & $<$ 12
% \end{tabular}
% \end{ruledtabular}
% \end{table}

\begin{figure}
\includegraphics[width=\columnwidth]{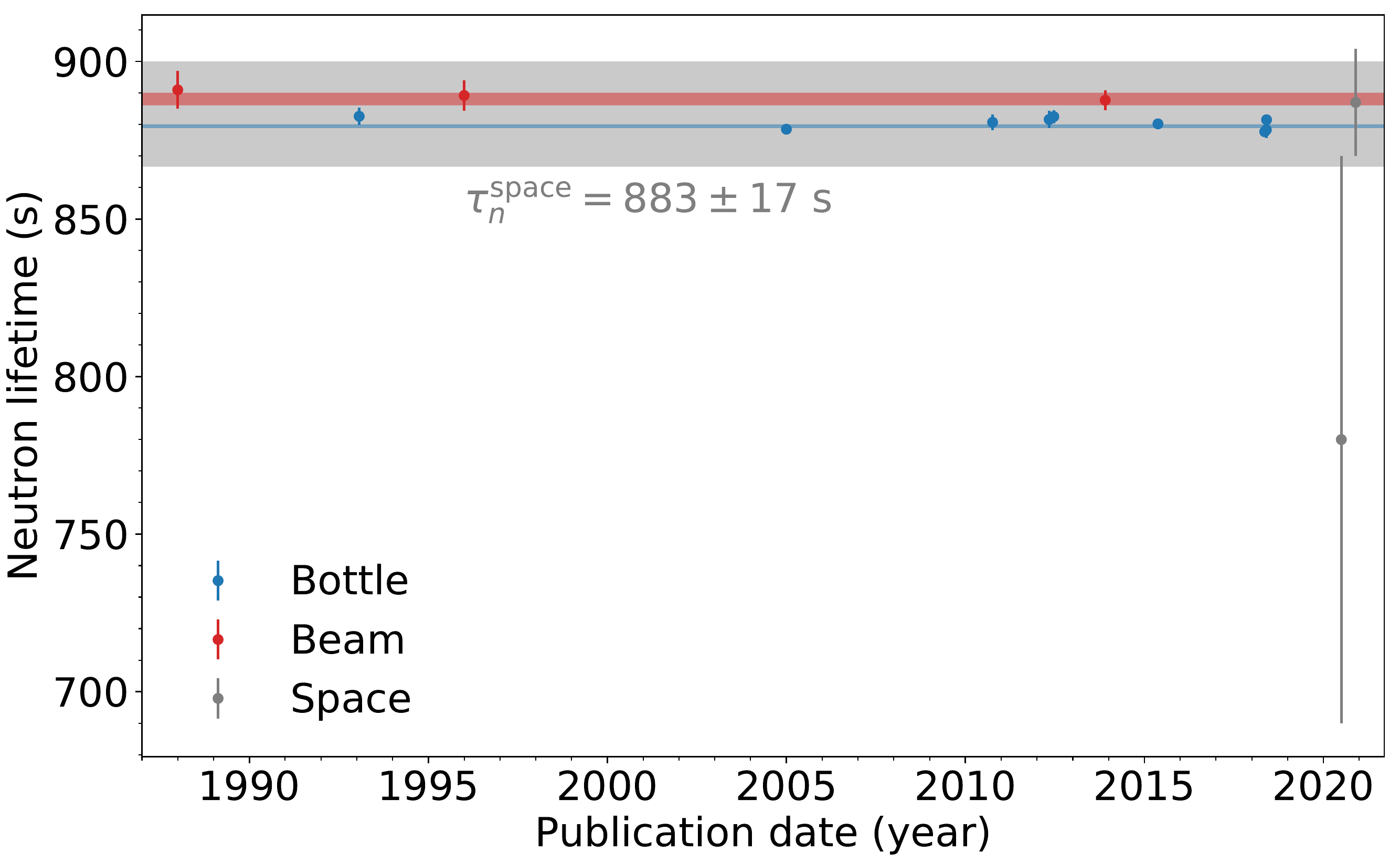}
\caption{\label{fig:history} Particle Data Group and other recent measurements of $\tau_n$ \citep{PDG2020,Yue2013,Byrne1996,Spivak1988}, along with those of this paper and \citet{Wilson2020}. The shaded regions represent the standard error on the uncertainty-weighted mean lifetime in each class of measurement.}
\end{figure}

% The results summarized in Table~\ref{tab:sys} and the data-to-model comparison presented in Fig.~\ref{fig:CS}(c) show 
We find $\tau_n=887 \pm 14_\text{stat}{\:^{+7}_{-3\:\text{syst}}}$~s, which is a 0.5$\sigma$ difference from the Particle Data Group (PDG) value of $879\pm0.4$~s \citep{PDG2020}. This result is shown in comparison with laboratory-based results and the previous space based result from \citet{Wilson2020} in Fig.~\ref{fig:history}. The measurement presented here further demonstrates the feasibility of measuring $\tau_n$ using a space-based experiment.  Additionally, the extension of the modeling to include non-uniform surface compositions extends the range of bodies where the space-based $\tau_n$ measurement might be performed and reduces what was a major systematic uncertainty in the earlier space-based $\tau_n$ measurement. The combined space-based measurement with standard error (shown as a grey region in Fig.~\ref{fig:history}) of $\tau_n^\textrm{space}=883\pm17$~s is in good agreement with the PDG value. 

\section{\label{sec:Conclusions}Conclusions}
Using data taken by the neutron detectors on NASA's Lunar Prospector spacecraft during the elliptical orbits completed immediately following the spacecraft's capture into a lunar orbit we found the neutron lifetime to be $\tau_n=887 \pm 14_\text{stat}{\:^{+7}_{-3\:\text{syst}}}$~s.  In combination with the previous measurement of $\tau_n$ from space using MESSENGER data \citep{Wilson2020}, this result firmly establishes the feasibility of making a measurement of $\tau_n$ from space. The statistical uncertainties are large compared with laboratory measurements due to the relatively short duration of the elliptical phase of the LP mission. Consequently, small amounts of data are taken, which is a result of the LP mission not being designed for this analysis. However, the modeling developed in this study, which enables non-uniform surfaces to be simulated, provides a key component to carrying out more thorough measurements of $\tau_n$ around airless bodies. Since the space-based method of constraining $\tau_n$ has entirely separate systematic uncertainties to the two existing classes of laboratory experiments, future space-based measurements with higher statistical precision and well-constrained systematics may provide a route to make progress beyond the current disagreement between the bottle and beam results. Achieving this goal would require a reduction of the systematics to the 1~s level, which necessitates further development of the space-based method that builds on the result of this paper.

\begin{acknowledgments}
We wish to thank the anonymous reviewer whose comments spurred substantial improvement in this paper. We gratefully acknowledge the support of the U.S. Department of Energy Office of Nuclear Physics (Grant No. DE-SC0019343). VRE was supported by the Science and Technology Facilities Council (STFC) grant ST/P000541/1. JAK acknowledges support from STFC grants ST/N001494/1 and ST/T002565/1. 
\end{acknowledgments}

\appendix
\section{\label{sec:comps}Model Compositions}

\begin{table*}
\caption{\label{tab:comps}The elemental mass fractions of the major crustal terranes used in the reference model: the Procellarum KREEP terrane (PKT), non-PKT maria (nPKT), South Pole Aitken terrane (SPAT),  Feldspathic highland terrane (FHT), and pure anorthosite terrane (PAN) \citep{Jolliff2000,Peplowski2016}.}
\begin{ruledtabular}
\begin{tabular}{l*{4}{r}r}\\
Element & FHT & PAN & SPA & nPKT & PKT \\
 \hline  
 H  & 5.13$\times 10^{-5}$ & 2.86$\times 10^{-5}$ & 5.37$\times 10^{-5}$ & 4.35$\times 10^{-5}$ & 4.91$\times 10^{-5}$  \\
 O  &    0.455 &    0.463 &    0.444 &    0.418 &    0.403  \\
 Na &  0.00228 &  0.00236 &  0.00225 &  0.00235 &  0.00227  \\
 Mg &   0.0368 &   0.0271 &   0.0489 &   0.0578 &   0.0562  \\
 Al &    0.135 &    0.146 &    0.114 &   0.0695 &    0.075  \\
 Si &    0.214 &    0.221 &    0.209 &    0.198 &    0.186  \\
 K  & 0.000441 & 0.000306 &  0.00101 &   0.0012 &  0.00254  \\
 Ca &     0.12 &    0.125 &    0.115 &    0.104 &    0.114  \\
 Ti &  0.00278 &  0.00216 &  0.00539 &   0.0208 &   0.0204  \\
 Fe &   0.0344 &   0.0125 &   0.0596 &    0.128 &    0.141  \\
 Sm & 8.43$\times 10^{-7}$ &        0 & 3.32$\times 10^{-6}$ &  1.8$\times 10^{-6}$ & 7.86$\times 10^{-6}$  \\
 Gd & 9.87$\times 10^{-7}$ &        0 & 3.89$\times 10^{-6}$ & 2.11$\times 10^{-6}$ & 9.19$\times 10^{-6}$  \\
 Th & 8.69$\times 10^{-8}$ & 8.83$\times 10^{-8}$ & 2.23$\times 10^{-6}$ & 2.72$\times 10^{-6}$ & 6.61$\times 10^{-6}$  \\
 U  & 2.36$\times 10^{-8}$ & 2.40$\times 10^{-8}$ & 6.06$\times 10^{-7}$ & 7.41$\times 10^{-7}$ & 1.80$\times 10^{-6}$
\end{tabular}
\end{ruledtabular}
\end{table*}

\begin{table*}
\caption{\label{tab:comps2}The elemental mass fractions of the major crustal terranes used in the perturbed, `1$\sigma$', models: the Procellarum KREEP terrane (PKT), non-PKT maria (nPKT), South Pole Aitken Terrane (SPAT),  Feldspathic Highland Terrane (FHT), and Pure Anorthosite terrane (PAN) \citep{Jolliff2000,Peplowski2016}.}
\begin{ruledtabular}
\begin{tabular}{l*{4}{r}r} \multicolumn{6}{c}{$\bf{+1\sigma}$}\\
\hline  
Element & FHT & PAN & SPA & nPKT & PKT \\
\hline  
 H  & 5.21$\times 10^{-5}$ & 2.92$\times 10^{-5}$ & 5.44$\times 10^{-5}$ & 4.39$\times 10^{-5}$ & 4.93$\times 10^{-5}$  \\
 O  &     0.46 &    0.466 &    0.452 &     0.43 &    0.417  \\
 Na &  0.00232 &  0.00241 &  0.00228 &  0.00237 &  0.00229  \\
 Mg &   0.0373 &   0.0277 &   0.0495 &   0.0584 &   0.0566  \\
 Al &    0.122 &    0.132 &    0.101 &   0.0551 &   0.0613  \\
 Si &    0.218 &    0.228 &    0.212 &    0.199 &    0.186  \\
 K  & 0.000454 &  0.00032 &  0.00103 &  0.00122 &  0.00257  \\
 Ca &    0.122 &    0.128 &    0.117 &    0.105 &    0.114  \\
 Ti &   0.0031 &  0.00241 &  0.00604 &   0.0222 &    0.022  \\
 Fe &   0.0348 &    0.013 &   0.0592 &    0.127 &    0.139  \\
 Sm & 8.56$\times 10^{-7}$ &        0 & 3.37$\times 10^{-6}$ & 1.82$\times 10^{-6}$ &  7.9$\times 10^{-6}$  \\
 Gd & 1.00$\times 10^{-6}$ &        0 & 3.94$\times 10^{-6}$ & 2.13$\times 10^{-6}$ & 9.24$\times 10^{-6}$  \\
 Th & 8.95$\times 10^{-8}$ & 9.13$\times 10^{-8}$ & 2.26$\times 10^{-6}$ & 2.74$\times 10^{-6}$ & 6.64$\times 10^{-6}$  \\
 U  & 2.78$\times 10^{-8}$ & 2.83$\times 10^{-8}$ & 6.51$\times 10^{-7}$ & 7.88$\times 10^{-7}$ & 1.85$\times 10^{-6}$ \\
 \hline  
\multicolumn{6}{c}{$\bf{-1\sigma}$}\\
 \hline  
 Element & FHT & PAN & SPA & nPKT & PKT \\
 \hline  
 H  & 5.06$\times 10^{-5}$ &  2.8$\times 10^{-5}$ & 5.31$\times 10^{-5}$ & 4.31$\times 10^{-5}$ & 4.88$\times 10^{-5}$ \\
 O  &     0.45 &     0.46 &    0.437 &    0.406 &     0.39 \\
 Na &  0.00225 &   0.0023 &  0.00222 &  0.00233 &  0.00226 \\
 Mg &   0.0362 &   0.0265 &   0.0483 &   0.0573 &   0.0559 \\
 Al &    0.147 &     0.16 &    0.126 &   0.0836 &   0.0885 \\
 Si &    0.209 &    0.214 &    0.206 &    0.198 &    0.186 \\
 K  & 0.000427 & 0.000293 & 0.000988 &  0.00118 &  0.00252 \\
 Ca &    0.118 &    0.122 &    0.114 &    0.103 &    0.113 \\
 Ti &  0.00246 &  0.00192 &  0.00474 &   0.0194 &   0.0188 \\
 Fe &    0.034 &   0.0121 &   0.0599 &     0.13 &    0.143 \\
 Sm & 8.31$\times 10^{-7}$ &        0 & 3.28$\times 10^{-6}$ & 1.79$\times 10^{-6}$ & 7.81$\times 10^{-6}$ \\
 Gd & 9.72$\times 10^{-7}$ &        0 & 3.84$\times 10^{-6}$ & 2.09$\times 10^{-6}$ & 9.14$\times 10^{-6}$ \\
 Th & 8.43$\times 10^{-8}$ & 8.53$\times 10^{-8}$ & 2.19$\times 10^{-6}$ &  2.7$\times 10^{-6}$ & 6.58$\times 10^{-6}$ \\
 U  & 1.96$\times 10^{-8}$ & 1.98$\times 10^{-8}$ & 5.63$\times 10^{-7}$ & 6.95$\times 10^{-7}$ & 1.76$\times 10^{-6}$
\end{tabular}
\end{ruledtabular}
\end{table*}
% \begin{sidewaystable}
% \caption{\label{tab:comps}The composition of the major crustal terranes defined in the model of the lunar neutron flux}
% \begin{ruledtabular}
% \begin{tabular}{l*{12}{c}r}\\
% Name & H & O & Na & Mg & Al & Si & K & Ca & Ti & Mn & Fe & Sm & Gd\\
% \hline
% PKT  & 0.4252 & 0.0035 & 0.0478 & 0.0666 & 0.1963 & 0.0000 & 0.0839 & 0.0476 & 0.0016 & 0.1275 & 1.38$\times10^{-5}$ & 1.80$\times10^{-5}$ \\
% nPKT  & 0.4401 & 0.0029 & 0.0584 & 0.1204 & 0.2113 & 0.0000 & 0.1051 & 0.0030 & 0.0008 & 0.0580 & 3.20$\times10^{-6}$ & 4.40$\times10^{-6}$ \\
% SPA  & 0.4599 & 0.0035 & 0.0362 & 0.1442 & 0.2098 & 0.0000 & 0.1041 & 0.0032 & 0.0005 & 0.0387 & 0.0000 & 0.0000 \\
% FHT  & 0.4560 & 0.0045 & 0.0051 & 0.1763 & 0.2066 & 0.0000 & 0.1359 & 0.0008 & 0.0000 & 0.0148 & 0.0000 & 0.0000 \\
% PAN  & 0.4682 & 0.0029 & 0.0016 & 0.1864 & 0.2012 & 0.0003 & 0.1352 & 0.0001 & 0.0002 & 0.0039 & 0.0000 & 0.0000
% \end{tabular}
% \end{ruledtabular}
% \end{sidewaystable}
% \bibliography{nl}{}
% \bibliographystyle{apsrev4-2}

%

\end{document}